\documentclass[a4paper,11pt,reqno]{amsart}

\usepackage{hyperref}
\usepackage{amssymb}
\usepackage{eucal}
\usepackage{eufrak}
\usepackage{graphics}
\usepackage{graphicx}
\usepackage{epsfig}
\usepackage{multicol}
\usepackage{xypic}
\usepackage{amsmath}
\usepackage{wasysym}
\usepackage{dsfont}
\usepackage{xypic}
\usepackage{amsmath}

\newtheorem{theorem}{Theorem}[section]
\newtheorem{proposition}{Proposition}[section]
\newtheorem{definition}{Definition}[section]

\newtheorem{remark}[definition]{Remark}
\newtheorem{example}[definition]{Example}

\newcommand{\se}{\mathfrak{se}(3)}

\newcommand{\Om}{\Omega}
\newcommand{\so}{\mathfrak{so}(3)}
\newcommand{\dso}{\mathfrak{so}(3)^{*}}
\newcommand{\pot}{\mathcal{W}}

\newcommand{\Flder}{\rightarrow}
\newcommand{\lp}{\left(}
\newcommand{\rp}{\right)}
\newcommand{\lc}{\left\{}
\newcommand{\rc}{\right\}}
\newcommand{\der}{\partial}

\newcommand{\bra}{\langle}
\newcommand{\ket}{\rangle}
\newcommand{\R}{\mathds{R}}      
\newcommand{\F}{\mathds{F}}

\usepackage{amssymb}

\newcommand{\proa}{A^*G \mbox{$\;$}_{\tau^*} \kern-3pt\times_\alpha
G \mbox{$\;$}_\beta \kern-3pt\times_{\tau^*} A^*G}

\newcommand{\alg}{\mathfrak{so}(3)}

\newcommand{\e}{\mbox{exp}}
\newcommand{\Ad}{\mbox{Ad}}
\newcommand{\ca}{\mbox{cay}}
\newcommand{\ad}{\mbox{ad}}

\newcommand{\potint}{\mathcal{W}^{int}}
\newcommand{\potext}{\mathcal{W}^{ext}}
\newcommand{\potbar}{W^{int}}
\newcommand{\Su}{\Sigma_{u}}
\newcommand{\Sv}{\Sigma_{v}}
\newcommand{\su}{\lp\Sigma_{u}\rp}
\newcommand{\sv}{\lp\Sigma_{v}\rp}
\newcommand{\caso}{\mbox{cay}_{SO(3)}}

\newcommand{\Upso}{\Upsilon^{SO(3)}}
\newcommand{\Uptres}{\Upsilon^{\R^{3}}}

\begin{document}


%

%

\title[Discrete Euler-Poincar\'e equations and optimal control]{Discrete second-order Euler-Poincar\'e equations. Applications to optimal control
}

\author{Leonardo Colombo}
\address{L.Colombo: Instituto de Ciencias Matem\'aticas (CSIC-UAM-UC3M-UCM), Calle Nicol\'as Cabrera 15, 28049 Madrid, Spain} \email{leo.colombo@icmat.es}

\author{Fernando Jim\'enez}
\address{F.Jim\'enez: Instituto de Ciencias Matem\'aticas (CSIC-UAM-UC3M-UCM), Calle Nicol\'as Cabrera 15, 28049 Madrid, Spain} \email{fernando.jimenez@icmat.es}

\author{David Mart\'in de Diego}
\address{D.\ Mart\'{\i}n de Diego: Instituto de Ciencias Matem\'aticas (CSIC-UAM-UC3M-UCM), Calle Nicol\'as Cabrera 15, 28049 Madrid, Spain} \email{david.martin@icmat.es}

\maketitle

\begin{abstract}
In this paper we will discuss some new developments in the design of
numerical methods for optimal control problems of
Lagrangian systems on Lie groups. We will construct these geometric
integrators using discrete variational calculus on Lie groups,
deriving a discrete version of the second-order Euler-Lagrange
equations. Interesting applications as, for instance, a discrete
derivation of the Euler-Poincar\'e equations for second-order
Lagrangians and its application to optimal control of a rigid body,
and of a Cosserat rod are shown at the end of the paper.

\end{abstract}


\section{Introduction}

The goal of this paper is to study, from a geometric point of view,
variational integrators for optimal control problems of mechanical
systems defined on finite dimensional Lie groups, and its
applications in optimal control theory. Our motivation is the
control of autonomous vehicles modeled as rigid bodies (as an
evolution equation in time).

We use the theory of discrete mechanics based on discrete
variational calculus \cite{mawest}. In particular, we use Hamilton's
principle yielding the set of discrete paths that approximately
satisfy the dynamics. This is achieved by formulating a second order
discrete variational problem solved through discrete Hamilton's
principle on Lie groups and obtaining a variational numeric
integrator. Such formulation gives us the preservation of important
geometric properties of the mechanical system, such as momentum,
symplecticity, group structure, good behavior of the energy, etc
\cite{Hair}.


A typical  optimal control problem consists on finding a trajectory
of the state variables and controls $(g(t),\xi(t),u(t))$ given fixed
initial and final conditions $(g(0),\xi(0))$ and $(g(T),\xi(T))$
respectively, and, as well, minimizing the cost functional defined
by $$J(u,T)=\int_{0}^{T}\|u(t)\|^{2}dt;$$ here, $g(t)$ evolves on a
Lie group $G$, $\xi(t)$ on the associated Lie algebra ${\mathfrak
g}$ and $u(t)$ on the space of admissible controls.

Our approach is based on recently developed structure-preserving
numerics integrators for optimal control problems (see
\cite{CoMaZu210},\cite{CoMaZu2011},\cite{Marin}, \cite{KM},
\cite{TL}, \cite{objuma} and references therein) based on solving a
discrete optimal control problem as a discrete higher-order
variational problem with higher-order constraints (see \cite{Bl} for
the continuous case) which are used for simulating and controlling
the dynamics for satellites, spacecrafts, underwater vehicles,
mobile robots, helicopters, wheeled vehicles, mobile robots, etc
\cite{bullolewis}.

\subsection{Background: Discrete Mechanics and variational integrators}
\label{section2} Let $Q$ be a $n$-dimensional differentiable
manifold, the configuration manifold,  with local coordinates
$(q^i)$, $1\leq i\leq n$. Denote by $TQ$ its tangent bundle with
induced coordinates $(q^i, \dot{q}^i)$. Given a Lagrangian function
$L:TQ\rightarrow \R$, the Euler-Lagrange equations are
\begin{equation}\label{qwer}
\frac{d}{dt}\left(\frac{\partial L}{\partial\dot
q^i}\right)-\frac{\partial L}{\partial q^i}=0, \quad 1\leq i\leq n.
\end{equation}
These equations are a system of implicit second order differential
equations.

 In the sequel, we will assume that the
Lagrangian is \textbf{regular}, that is, the matrix
$\left(\frac{\partial L}{\partial \dot q^i
\partial \dot q^j}\right)$ is non-singular. It is well known that
the origin of these equations is variational (see
\cite{AbMa},\cite{MaRa}).

Variational integrators \cite{mawest} are derived from a discrete
variational principle. These integrators also retain some of main
geometric properties of the continuous system, such as
simplecticity, momentum conservation and a good behavior of the
energy associated with the Lagrangian system (see \cite{Hair} and
references therein).

In the sequel we will review the construction of this type of
geometric integrators.

A \textbf{discrete Lagrangian} is a map $L_d\colon Q \times Q\to
\R$, which may be considered as an approximation of the integral
action defined by a continuous  Lagrangian $L\colon TQ\to \R,$
\[
L_d(q_0, q_1)\approx \int^h_0 L(q(t), \dot{q}(t))\; dt
\]
where $q(t)$ is a solution of the Euler-Lagrange equations for $L$;
$q(0)=q_0$, $q(h)=q_1$ and the time step $h>0$ is small enough.

 Define the \textbf{action sum} $\mathcal{A}_d\colon Q^{N+1}\to \R$,   corresponding to the
Lagrangian $L_d$ by
$$
{\mathcal{A}_d}=\sum_{k=1}^{N}  L_d(q_{k-1}, q_{k}),
$$
where $q_k\in Q$ for $0\leq k\leq N$, where $N$ is the number of
steps. The discrete variational principle then requires that $\delta
{\mathcal{A}_d}=0$ where the variations are taken with respect to
each point $q_k,$ $1\leq k\leq N-1$ along the path, and the
resulting equations of motion (system of difference equations) given
fixed endpoints $q_0$ and $q_N,$ are
\begin{equation}\label{discreteeq}
 D_1L_d( q_k, q_{k+1})+D_2L_d( q_{k-1}, q_{k})=0,
\end{equation}
where $D_1$ and $D_2$ denote the derivative to the Lagrangian
respect the first and second arguments, respectively.

These  equations are usually called \textbf{discrete Euler--Lagrange
equations}. Under some regularity hypotheses (the matrix
$(D_{12}L_d(q_k, q_{k+1}))$ is regular), it is possible to define a
(local) discrete flow $ \Upsilon_{L_d}\colon Q\times Q\to Q\times
Q$, by $\Upsilon_{L_d}(q_{k-1}, q_k)=(q_k, q_{k+1})$ from
(\ref{discreteeq}).

 We introduce now the two discrete Legendre transformations associated to $L_{d}$:
\begin{eqnarray}\nonumber\label{TransLegDiscre}
{\F}^{-}L_{d}:Q\times Q &\rightarrow& T^{*}Q\\\nonumber
\lp q_{0},q_{1}\rp&\mapsto&\lp q_{0},-D_{1}L_{d}\lp q_{0},q_{1}\rp\rp,\\\\\nonumber
{\F}^{+}L_{d}:Q\times Q &\rightarrow& T^{*}Q\\\nonumber
\lp q_{0},q_{1}\rp&\mapsto&\lp q_{1},D_{2}L_{d}\lp q_{0},q_{1}\rp\rp,
\end{eqnarray}
and the discrete Poincar\'e-Cartan
2-form $\omega_{d}=\lp{\F}^{+}L_{d}\rp^{*}\omega_{Q}=\lp{\F}^{-}L_{d}\rp^{*}\omega_{Q}$, where $\omega_{Q}$
is the canonical symplectic form on $T^{*}Q$. $\omega_d$ is a symplectic form if the discrete Lagrangian is regular, which is
indeed equivalent to  ${\F}^{-}L_{d}$ (or ${\F}^{+}L_{d}$) being a local diffeomorphism.

 The discrete algorithm determined by $\Upsilon_{L_d}$ preserves the
(pre-)symplectic form on $T^{*}(Q\times Q),$ $\omega_d$, i.e.,
$\Upsilon_{L_d}^*\omega_d=\omega_d$. Moreover, if the discrete
Lagrangian is invariant under the diagonal action of a Lie group
$G$, then the discrete momentum map $J_d\colon Q\times Q \to
\mathfrak{g}^*$ defined by
\[ \langle
J_d(q_k, q_{k+1}), \xi\rangle=\langle D_2L_d(q_k, q_{k+1}),
\xi_Q(q_{k+1})\rangle \] is preserved by the discrete flow.
Therefore, these integrators are symplectic-momentum preserving.
Here, $\xi_Q$ denotes the fundamental vector field determined by
$\xi\in \mathfrak{g}$, where $\mathfrak{g}$ is the Lie algebra of
$G$, $$\xi_{Q}(q)=\frac{d}{dt}\Big|_{t=0} (exp(t\xi)\cdot q)$$ for
$q\in Q$ (see \cite{mawest} for more details).

\begin{example}
{\rm For instance we consider a Lagrangian
$L(q,\dot{q})=\frac{1}{2}\dot{q}^{T}M\dot{q}-V(q),$ where
$q\in\R^{3}$, $M$ being a symmetric non-degenerate matrix and $V$ a potential function. From this
Lagrangian we construct the discrete Lagrangian taking an Euler's discretization,
\[L_d(q_k,q_{k+1})=
h\left[\left(\frac{q_{k+1}-q_{k}}{h}\right)^{T}M\left(\frac{q_{k+1}-q_{k}}{h}\right)-V(q_k)\right].\]
We compute $D_1\,L_d$ and $D_2\,L_d\;$:
\begin{eqnarray*}
D_1L_d(q_k,q_{k+1})&=&-M\left(\frac{q_{k+1}-q_k}{h}\right)-h\nabla
V(q_k),\\
D_2L_d(q_{k-1},q_k)&=&M\left(\frac{q_k-q_{k-1}}{h}\right),
\end{eqnarray*}
which leads to the discrete Euler-Lagrange equations:
\[M\left(\frac{q_{k+1}-2q_k+q_{k-1}}{h^2}\right)=-\nabla V(q_k).\]
We observe that these equations give rise a natural discrete version
of the Newton's law $M\ddot{q}=-\nabla V(q)$, using a simple finite
difference rule for the derivative (see \cite{mawest}).
}

\end{example}

\subsection{Organization of the paper}

The paper is structured as follows. In Section \ref{section3} we
recall some results given in \cite{MaPeSh} about Hamilton's
principle on Lie groups and the discrete Euler-Poincar\'e equations.
The new proposed method appears in Section \ref{section4}. First, we
derive the continuous second-order Euler-Poincar\'e equations on Lie
groups from Hamilton's principle; next, we construct from a
discretization of the  Lagrangian and through  discrete variational
calculus the discrete second-order Euler-Lagrange and
Euler-Poincar\'e equations. The discrete higher-order Euler-Lagrange
and discrete higher order Euler-Poincar\'e equations are derived
using discrete Hamilton's principle. In the last section, we apply
these techniques to optimal control of mechanical systems and we
analyze two examples of optimal control on a rigid body on the Lie
group $SO(3)$ and on a Cosserat rod defined on $SE(3)$.

\section{Discrete mechanics on Lie groups}
\label{section3}


In this
section we recall the discrete mechanics on Lie groups and
Hamilton's principle on Lie groups for the formulation of
Euler-Poincar\'e equations.

\subsection{Discrete Hamilton's principle on Lie groups and Euler-Poincar\'e equations}
\label{Ref}
If the configuration space is a Lie group $G$, then the discrete
trajectory is represented numerically using a set of $N +1$ points
$(g_0, g_1, \ldots, g_N)$ with $g_i\in G$, $0\leq i\leq N$.

A way to  discretize a continuous problem is using a \emph{retraction map} $\tau: {\mathfrak g}\to G$ which is an analytic local diffeomorphism which maps a neighborhood of $0\in {\mathfrak g}$ to a neighborhood of the neutral element $e\in G$.
As a consequence, it is possible to deduce that $\tau(\xi)\tau(-\xi)=e$ for all $\xi \in \mathfrak{g}$.
The retraction map  is used
to express small discrete changes in the group configuration
through unique Lie algebra elements (see \cite{KM}), namely $\xi_k=\tau^{-1}(g_{k}^{-1}g_{k+1})/h$, where $\xi_k\in\mathfrak{g}$. That is, if $\xi_{k}$ were regarded as an average velocity between $g_{k}$ and $g_{k+1}$, then $\tau$ is an approximation to the integral flow of the dynamics. The difference $g_{k}^{-1}\,g_{k+1}\in G$, which is an element of a nonlinear space, can now be represented by the vector $\xi_{k},$ in order to enable unconstrained optimization in the linear space $\mathfrak{g}$ for optimal control purposes.

It will be useful in the sequel, mainly in the derivation of the discrete equations of motion, to define the {\it right trivialized} tangent retraction map as
\[
T_{\xi}\,\tau=T_e r_{\tau_{\xi}}\circ \hbox{d}\tau_{\xi}.
\]
Useful and complementary definition of the right trivialized (and its inverse) is the following (\cite{MunteKaas}, \cite{Rabee}):

\begin{proposition}\label{Retr}
{\rm Given a map $\tau:\mathfrak{g}\Flder G$, its right trivialized tangent $\mbox{d}\tau_{\xi}:\mathfrak{g}\Flder\mathfrak{g}$ and its inverse $\mbox{d}\tau_{\xi}^{-1}:\mathfrak{g}\Flder\mathfrak{g}$, are such that for $g=\tau(\xi)\in G$ and $\eta\in\mathfrak{g}$, the following holds}
\begin{eqnarray*}
&&\der_{\xi}\tau(\xi)\,\eta=\mbox{d}\tau_{\xi}\,\eta\,\tau(\xi),\\
&&\der_{\xi}\tau^{-1}(g)\,\eta=\mbox{d}\tau^{-1}_{\xi}(\eta\,\tau(-\xi)).
\end{eqnarray*}
\end{proposition}

 An example of retraction map is the exponential
map at the identity $e$ of the group $G,$
$exp_{e}:\mathfrak{g}\rightarrow G$. We recall that for a finite dimensional Lie grup, $exp_e$ is
locally a diffeomorphism and gives rise a natural chart
\cite{MaPeSh}. Then, there exists a neighborhood $U$ of $e\in G$
such that $exp_e^{-1}:U\rightarrow exp_e^{-1}(U)$ is a local
$\mathcal{C}^{\infty}$ diffeomorphism. A chart at $g\in G$ is given
by $\Psi_{g}=exp_{e}^{-1}\circ l_{g^{-1}},$ where $l$ denote the
left-translation of an element of the group.
%

In general, it is not easy to work with the exponential. For instance, if we are considering matrix groups, the right trivialized derivative and its inverse are defined by infinite series
\begin{eqnarray*}
\mbox{d}\e_{x}\,y&=&\sum_{j=0}^{\infty}\frac{1}{(j+1)!}\,\ad_{x}^{j}\, y,\\
\mbox{d}\e_{x}^{-1}\,y&=&\sum_{j=0}^{\infty}\frac{B_{j}}{j!}\,\ad_{x}^{j}\, y,
\end{eqnarray*}
where $B_{j}$ are the Bernoulli numbers, $x,y\in\mathfrak{g}$ and $\ad_x\,y=[x,y]$ is the usual matrix bracket (see \cite{Hair}). Tipically, these expressions are truncated in order to achieve a desired order of accuracy.

    In consequence
it will be useful to use a different retraction map. More concretely, the Cayley map (see \cite{Hair} for further details) will provide us a proper framework in the examples shown below.

The following theorem, regardless of the {\it retraction structure} locally relating $G$ and $\mathfrak{g}$, gives us the relation between the discrete
Euler-Lagrange equations and the discrete Euler-Poincar\'e
equations.

\begin{theorem}\cite{Mars3}
Let $G$ be a Lie group and $L_d:G\times G\rightarrow\mathbb{R}$ a
discrete Lagrangian function. We suppose that $L_d$ is
left-invariant over the diagonal action $(i.e;
L_d(gg_k,gg_{k+1})=L_{d}(g_k,g_{k+1}) \hbox{ with } g\in G )$. Let
$\tilde{l}_d:G\rightarrow\mathbb{R}$ be the restriction to the
identity (that is, $\tilde{l}_d:(G\times G)/ G\simeq
G\rightarrow\mathbb{R},$
$\tilde{l}_d(g_{k}^{-1}g_{k+1})=L_d(g_k,g_{k+1})$). For a pair of
points $(g_k,g_{k+1})\in G\times G,$ we consider
$W_k=g^{-1}_{k}g_{k+1}$ (where $g_{k}^{-1}=i(g_k), i:G\rightarrow G$
the inversion map of the Lie group $G$). Then the following
assertions are equivalent:
\begin{enumerate}
\item $(g_k)_{0\leq k\leq N}$ satisfies the discrete Euler-Lagrange equations for $L_d$.
\item $(g_k)_{0\leq k\leq N}$ extremize the discrete action $$(g_k)_{0\leq k\leq N}\mapsto\sum_{k=0}^{N-1}L_d(g_k,g_{k+1})$$ for all variation with initial and final fixed points.
\item The discrete Euler-Poincar\'e equations $$r^{*}_{W_k} \tilde{l}'_d(W_k)-l_{W_{k-1}}^{*}\tilde{l}'_d(W_{k-1})=0 \qquad k=1,\ldots,N$$
    hold, where $l$ and $r$ are the left- and right-translation of
    the Lie group and $'$denote the partial derivative.
\item $(W_k)_{0\leq k\leq N-1}$ extremize $$(W_k)_{0\leq k\leq N-1}\mapsto\sum_{k=0}^{N-1}\tilde{l}_d(W_k)$$ for all variations $\delta W_k=-\Sigma_k W_k + W_k\Sigma_{k+1}$ with $\Sigma_0=\Sigma_N=0;$ where
$\Sigma_k\in\mathfrak{g}$ is given by $\Sigma_k=g_k\delta g_k.$
\end{enumerate}
\end{theorem}

\section{Continuous and discrete Euler-Poincar\'e equations for second order lagrangians}
\label{section4}

In this section we derive, from a variational point of view, the
discrete and continuous Euler-Lagrange equations for second-order
Lagrangians defined on Lie groups:
the second order Euler-Poincar\'e equations in the continuous and
discrete setting.

Consider a mechanical system  determined by a Lagrangian
$L:TG\longrightarrow \R.$ It is well known that the tangent bundle
$TG$ can be left-trivialized as $TG\simeq G\times\mathfrak{g}$,
where ${\mathfrak g}$ is the Lie algebra of a Lie group $G.$ The
motion of the mechanical system is described by applying the
following principle
\begin{equation}\label{ldp}
\delta \int^T_0 L(g(t), \xi(t))\, dt=0
\end{equation}
for all variations $\delta \xi(t)$ of the form $\delta
\xi(t)=\dot{\eta}(t)+[\xi(t), \eta(t)]$, where $\eta$ is an
arbitrary curve on the Lie algebra with $\eta(0)=0=\eta(T)$ and
$\delta g=g\eta$ (see \cite{MaRa}). This principle give rise to the
Euler-Lagrange equations \[\frac{d}{dt}\left( \frac{\delta L}{\delta
\xi}\right)=\hbox{ad}^*_{\xi}\left( \frac{\delta L}{\delta
\xi}\right)+l_g^*\frac{\delta L}{\delta g}\] where
$\hbox{ad}_{\xi}\eta=[\xi, \eta]$. If the Lagrangian $L$ is
left-invariant the above equations are written as
 \[\frac{d}{dt}\left( \frac{\delta L}{\delta
\xi}\right)=\hbox{ad}^*_{\xi}\left( \frac{\delta L}{\delta
\xi}\right)\] and are called the \textit{Euler-Poincar\'e equations}.
\subsection{Continuous setting}
In this subsection we deduce, from a variational principle, the
Euler-Poincar\'e equations for Lagrangians defined on
$T^{(2)}G\simeq G\times 2\mathfrak{g}$ from a left-trivization. One interesting application  of this theory will be the  optimal control of
mechanical systems as we will seen in the next section (see \cite{CMdD})

Let $L:T^{(2)}G\simeq G\times 2\mathfrak{g}\rightarrow\mathbb{R}$ be
a Lagrangian function, $L(g, \dot{g}, \ddot{g})\equiv L(g, \xi,
\dot{\xi})$ where  $\xi=g^{-1}\dot{g}$ (left-trivialization). The
problem consists on finding the critical curves of the functional
$$\mathcal{J}=\int_{0}^{T}L(g,\xi,\dot{\xi})dt$$ among all curves
satisfying the boundary conditions for arbitrary variations $\delta
g=\frac{d}{d\epsilon}\mid_{\epsilon=0}g_{\epsilon},$
where,$\epsilon\mapsto g_{\epsilon}$ is a smooth curve in $G$ such
that $g_0=g.$

We define, for any $\epsilon,$
$\xi_{\epsilon}:=g^{-1}_{\epsilon}\dot{g}_{\epsilon}.$ The
corresponding variations $\delta\xi$ induced by $\delta g$ are given
by $\delta\xi=\dot{\eta}+[\xi,\eta]$ where $\eta:=g^{-1}\delta
g\in\mathfrak{g}$ $(\delta g=g\eta)$. Therefore

\begin{eqnarray*}&&\delta\int_{0}^{T}L(g(t),\xi(t),\dot{\xi}(t))dt=\\
&&\frac{d}{d\epsilon}\Big|_{\epsilon=0}\int_{0}^{T}L(g_{\epsilon}(t),\xi_{\epsilon}(t),\dot{\xi}_{\epsilon}(t))dt=\\
&&\int_{0}^{T}\left(\langle \frac{\partial L}{\partial g},\delta g\rangle+\langle\frac{\delta L}{\delta\xi},\delta\xi\rangle+\langle\frac{\delta L}{\delta\dot{\xi}},\delta\dot{\xi}\rangle\right) dt=\\
 &&\int_{0}^{T}\left(\langle \frac{\partial L}{\partial g},\delta g\rangle+\langle\frac{\delta L}{\delta\xi},\delta\xi\rangle+\langle\frac{\delta L}{\delta\dot{\xi}},\frac{d}{dt}(\delta\xi)\rangle \right)dt=\\
&&\int_{0}^{T}\left(\langle \frac{\partial L}{\partial g}, \delta g\rangle+\langle\frac{\delta L}{\delta\xi},\delta\xi\rangle+\langle-\frac{d}{dt}\frac{\delta L}{\delta\dot{\xi}},\delta\xi\rangle\right) dt=\\
&&\int_{0}^{T}\left(\langle \frac{\partial L}{\partial g}, g\eta\rangle+\langle\frac{\delta L}{\delta\xi}-\frac{d}{dt}\frac{\delta L}{\delta\dot{\xi}},\frac{d}{dt}\eta+[\xi,\eta]\rangle\right) dt=\\
&&\int_{0}^{T}\Big\langle\left(-\frac{d}{dt}+ad^{*}_{\xi}\right)\left(\frac{\delta L}{\delta\xi}-\frac{d}{dt}\frac{\delta L}{\delta\dot{\xi}}\right),\eta\Big\rangle dt +\int_{0}^{T}\Big\langle l_{g}^{*}\left(\frac{\partial L}{\partial
g}\right),\eta\Big\rangle dt=0,
\end{eqnarray*}
where we have used integration by parts and the vanishing initial and endpoint conditions
$\eta(0)=\eta(T)=\dot{\eta}(0)=\dot{\eta}(T)=0$. Thus, the
stationary condition $\delta\mathcal{J}=0$ implies the
\textit{second-order Euler-Lagrange equations},
$$l_{g}^{*}\frac{\partial L}{\partial g} +
\left(-\frac{d}{dt}+ad^{*}_{\xi}\right)\left(\frac{\delta
L}{\delta\xi}-\frac{d}{dt}\frac{\delta
L}{\delta\dot{\xi}}\right)=0$$ that is, \begin{equation}\label{EP1}
l_{g}^{*}\frac{\partial L}{\partial g}-\frac{d}{dt}\frac{\delta
L}{\delta\xi}+\frac{d^2}{dt^2}\frac{\delta
L}{\delta\dot{\xi}}+ad^{*}_{\xi}\frac{\delta
L}{\delta\xi}-ad^{*}_{\xi}\left(\frac{d}{dt}\frac{\delta
L}{\delta\dot{\xi}}\right)=0.
\end{equation}

If the Lagrangian is invariant under an action of the Lie group, the
equations of motion are
\begin{equation}\label{EP2}
\frac{d^2}{dt^2}\frac{\delta
L}{\delta\dot{\xi}}-\frac{d}{dt}\frac{\delta
L}{\delta\xi}+ad^{*}_{\xi}\frac{\delta
L}{\delta\xi}-ad^{*}_{\xi}\left(\frac{d}{dt}\frac{\delta
L}{\delta\dot{\xi}}\right)=0.
\end{equation}
These equations are called \textit{second order Euler-Poincar\'e
equations.
}

In a recent paper \cite{FHR2010}, the authors studied invariant higher
order problems and obtain the equations (\ref{EP2}) working in a
reduced Lagrangian setting on $\mathfrak{g}\times\mathfrak{g}$.


\subsection{Discrete setting}

Now, we consider the associated discrete problem. The
second order tangent bundle is left-trivialized as $T^{(2)}G\simeq
G\times 2\mathfrak{g}$ and then we choose its natural discretization as
three copies of the Lie group (we recall that the prescribed
discretization of a Lie algebra $\mathfrak{g}$ is its associated Lie
group $G$). Consequently, we develop the discrete Euler-Lagrange
equations for the discrete Lagrangians defined on $G\times G\times
G=3G$.

Let $L_d: 3G\rightarrow\mathbb{R}$ be a discrete
Lagrangian where $G$ is a finite dimensional Lie group. As in the
previous section, we define $W_k=g_k^{-1}g_{k+1}$. Taking variations
for $W_k,$ where we denote $\Sigma_k=g_k^{-1}\delta g_k$, we obtain
\begin{eqnarray*}
\delta W_k&=&-g_k^{-1}\delta g_kg_{k}^{-1}g_{k+1} + g_k^{-1}\delta g_{k+1}\\
&=&-\Sigma_k W_k+ g_k^{-1}g_{k+1}g_{k+1}^{-1}\delta g_{k+1}\\
&=&-\Sigma_k W_k+ W_k\Sigma_{k+1},
\end{eqnarray*}
where $g_k,W_k\in G$ and $\Sigma_k\in\mathfrak{g}$.

The equations of motion are the critical paths of the discrete
action $$\sum_{k=0}^{N-2}L_d(g_k,W_k,W_{k+1})$$ with boundary
conditions $\Sigma_0=\Sigma_1=\Sigma_{N-1}=\Sigma_N=0$ since we are
assuming that $g_0$, $g_1$, $g_{N-1}$ and $g_N$ fixed.  Therefore,
after some computations we  obtain the equations

%
\begin{eqnarray*}
&&l_{g_{k-1}}^{*}D_1L_d(g_{k-1},W_{k-1},W_k)+l_{W_{k-1}}^{*}D_2L_d(g_{k-1},W_{k-1},W_{k})\\
&&-r_{W_k}^{*}D_2L_d(g_k,W_k,W_{k+1})-r_{W_k}^{*}D_3L_d(g_{k-1},W_{k-1},W_k)\\
&&+l_{W_{k-1}}^{*}D_3L_d(g_{k-2},W_{k-2},W_{k-1})=0
\end{eqnarray*}

These equation, together with the {\bf reconstruction equation}
$W_k=g_k^{-1}g_{k+1}$, are called \textit{discrete second order
Euler-Lagrange equations .}

If $L_d$ is $G$ invariant in the sense that $L_d(g_k, W_{k-1},
W_k)=L_d(hg_k,W_{k-1}, W_k)$ for all $h\in G$ then we can define the
reduced lagrangian $l_d: G\times G\to \R$ and the equations are
rewritten as
\begin{eqnarray*}
0&=&l_{W_{k-1}}^{*}D_1l_d(W_{k-1},W_{k})-r_{W_k}^{*}D_1l_d(W_k,W_{k+1})\\
&-&r_{W_k}^{*}D_2l_d(W_{k-1},W_k)+l_{W_{k-1}}^{*}D_2l_d(W_{k-2},W_{k-1})
\end{eqnarray*}
and are called the \textit{discrete second-order Euler-Poincar\'e
equations.}

\begin{remark}

{\rm
Is easy to extend these techniques for higher order discrete mechanics
(see \cite{benito}). Consider a mechanical system determined by a Lagrangian
$L:T^{(k)}G\longrightarrow \R.$ It is well known that the tangent
bundle $T^{(k)}G$ can be left-trivialized as
$T^{(k)}G\simeq G\times k\mathfrak{g}$, where ${\mathfrak g}$ is the
Lie algebra $G$.

Now, we consider the associated discrete problem. First, we replace
the higher order tangent bundle by $(k+1)$ copies of the group since the prescribed discretization of each $\mathfrak{g}$ is the Lie group $G$. At this point, we develop the discrete
Euler-Poincar\'e equations for the discrete Lagrangians defined on
$G\times kG$.

Let $L_d:G\times kG\rightarrow\mathbb{R}$ be a discrete Lagrangian
where $G$ is a finite dimensional Lie group. As before,  denote by
$W_i=g_i^{-1}g_{i+1}$ and
$\Sigma_i=g_i^{-1}\delta g_i$. Taking variations over $W_i$ we obtain
\begin{eqnarray*}
\delta W_i&=&-g_i^{-1}\delta g_i\,g_{i}^{-1}g_{i+1} + g_i^{-1}\delta g_{i+1}\\
&=&-\Sigma_i W_i+ g_i^{-1}g_{i+1}g_{i+1}^{-1}\delta g_{i+1}\\
&=&-\Sigma_i W_i+ W_i\Sigma_{i+1},
\end{eqnarray*}
where $g_i,W_i\in G$ and $\Sigma_i\in\mathfrak{g}$.

The equations of motion are the critical paths of the discrete action
$$\min\sum_{i=0}^{N-k}L_d(g_i,W_{(i,i+k-1)})$$ with boundary
conditions $\Sigma_0=\ldots=\Sigma_{k-1}=0,$
$\Sigma_{N-k+1}=\ldots=\Sigma_{N}=0$ and $g_0,\ldots, g_{k-1}$ and $g_{N-k+1},\ldots, g_{N}$
fixed.

Taking variations we deduce
\begin{eqnarray*}
&&\delta\sum_{i=0}^{N-k}L_d(g_i,W_{(i,i+k-1)})=\\
&&\sum_{i=k}^{N-k}\Big[D_1L_{d}(g_i,W_{(i,i+k)})\left(g_i\Sigma_i\right)
\\
&&+\sum_{j=2}^{k+1}D_jL_d(g_i,W_{(i,i+k-1)})\left(-\Sigma_{j+i-2}W_{j+i-2}+W_{j+i-2}\Sigma_{j+i-1}\right)\Big]
\end{eqnarray*}
where we denote by $W_{(i,j)}=(W_i, W_{i+1}, \ldots, W_{j-1}, W_j)$.

Therefore, the \textit{discrete higher-order Euler-Lagrange
equations} on $G\times kG$ are given by
\begin{eqnarray*}
&&0=l^{*}_{g_{i-1}}D_1L_d(g_{i-1},W_{(i-1,i+k-1)})\\
&&+\sum_{j=2}^{k+1}\left(l^{*}_{W_{i-1}}\right)D_jL_d(g_{i-j+1},W_{(i-j+1,i-j+k)})\\
&&-\sum_{j=2}^{k+1}\left(r^{*}_{W_{i}}\right)D_jL_d(g_{i-j+2},W_{(i-j+2,i-j+k+1)}).
\end{eqnarray*}
where $k\leq i\leq N-k$.

These equations, together with the reconstruction equation
$W_i=g_i^{-1}g_{i+1}$ are called the \textit{discrete higher-order
Euler-Lagrange equations}.
If $L_d$ is $G$-invariant, that is
$L_d(g_i,W_{(i,i+k-1)})=L_d(hg_i,W_{(i,i+k-1)})$ $\forall h\in G,$
we can consider the reduced Lagrangian
$l_d:kG\rightarrow\mathbb{R}.$ Then the \textit{discrete higher-order Euler-Poincar\'e equations} on the reduced space $kG$ are
given by

\begin{eqnarray*}
&&0=\sum_{j=2}^{k+1}\left(l^{*}_{W_{i-1}}\right)D_jL_d(W_{(i-j+1,i-j+k)})\\
&&-\sum_{j=2}^{k+1}\left(r^{*}_{W_{i}}\right)D_jL_d(W_{(i-j+2,i-j+k+1)}).
\end{eqnarray*}

}

\end{remark}

\section{Discrete Optimal control problems on Lie groups}
The proposal of this section is to study optimal control problems in the
case of fully actuated mechanical systems. The discrete
approximation to the solutions of the system have a purely discrete
variational formulation and as a consequence, the integrators defined
in this way are symplectic (Poisson)-momentum preserving. By using backward error analysis, it is well known that these integrators have a
good energy behavior (see \cite{mawest}).

As particular examples, we will  study the optimal control of the rigid body and the Cosserat rod. The configuration groups in these examples are $SO(3)$ and $SE(3)$ respectively. Both are particular cases of {\it quadratic} Lie groups, which are defined as
\[G=\lc Y\in GL(n, \R)\,\mid\,Y^{T}PY=Y\rc\]
where $P\in GL(n, \R)$ is a given matrix (here, $GL(n, \R)$ denotes the general linear
group of degree $n$). The corresponding Lie algebra is
\[\mathfrak{g}=\lc\Omega\in {\mathfrak gl}(n,
\R)\,\mid\,P\Omega+\Omega P=0\rc.\]
As mentioned in subsection \ref{Ref}, the Cayley map, defined for quadratic Lie groups as
\[
\ca(\xi)=\lp I-\frac{\xi}{2}\rp^{-1}\lp I+\frac{\xi}{2}\rp,
\]
where $\xi\in\mathfrak{g}$, also gives a useful and simpler discretization of these systems.

\subsection{Example: Rigid body}
\label{rb}
The rigid body problem is very well known in the literature. This
setting is deeply studied in \cite{KM,MeLee,MeLeeMc}
among other references.

The continuous equations of motion of the controlled rigid body
system are the following

\begin{eqnarray}\nonumber
\dot\Om_{(1)}&=&\rho_{1}\Om_{(2)}\Om_{(3)}+u_{1},\\\label{CuerpoRigido}
\dot\Om_{(2)}&=&\rho_{2}\Om_{(1)}\Om_{(3)}+u_{2},\\\nonumber
\dot\Om_{(3)}&=&\rho_{3}\Om_{(1)}\Om_{(2)}+u_{3},
\end{eqnarray}
where $(\Om_{(1)}, \Om_{(2)},\Om_{(3)})=\Omega$ and $(\dot\Om_{(1)},
\dot\Om_{(2)},\dot\Om_{(3)})=\dot\Om\in\R^3$, $u_{i}$ are the
control forces and $\rho_{i}\in\R$ are a redefinition of the inertia
momenta of the problem. In the sequel we will use the typical
identification of the Lie algebra of $SO(3),\mathfrak{so}(3)$ with
$\R^3$ by $\hat\cdot:\R^{3}\Flder\mathfrak{so}(3)$, that is if
$x=(x_{1},x_{2},x_{3})\in\R^{3}$
\[
\hat x=\lp\begin{array}{ccc}
0&-x_{3}&x_{2}\\
x_{3}&0&-x_{1}\\
-x_{2}&x_{1}&0
\end{array}
\rp\in\mathfrak{so}(3).
\]
Consequently $x\times y=-[\hat x,\hat y]=ad_{\hat x}\hat y$. With
some abuse of notation, we will directly identify $\R^{3}$ with
$\mathfrak{so}(3)$ by omitting the hat notation.

Our fixed boundary conditions are
$(R(0),\Om(0))$ and $(R(T),\Om(T))$,
where $R(t)\in SO(3)$ is the attitude of the rigid body subject to
the conditions $\dot R=R\Om$ and $\delta R=R\eta$, with $\eta$ an
arbitrary element of $\so$. Besides the equations, the cost
functional is
\[
\mathcal{C}=\int^{T}_{0}\frac{1}{2}u^{T}u\,dt,
\]
where $u=(u_{1},u_{2},u_{3})$. From eqs. (\ref{CuerpoRigido}) we can
work out $u$ in terms of $\Om$ and $\dot\Om$. Consequently, we can
define the function $l:\so\times\so\Flder\R$ in the following way
$$l(\Om,\dot\Om)=\frac{1}{2}u^{T}(\Om,\dot\Om)u(\Om,\dot\Om).$$
Therefore, the Lagrangian function has the following form:
\begin{eqnarray}\nonumber
l(\Om,
\dot\Om)&=&\frac{1}{2}\lp\dot\Om_{(1)}-\rho_{1}\Om_{(2)}\Om_{(3)}\rp^{2}+\frac{1}{2}\lp\dot\Om_{(2)}-\rho_{2}\Om_{(1)}\Om_{(3)}\rp^{2}+\\\label{LagCont}
&+&\frac{1}{2}\lp\dot\Om_{(3)}-\rho_{3}\Om_{(1)}\Om_{(2)}\rp^{2}.
\end{eqnarray}
With this redefinition, the cost functional becomes
\[
\mathcal{C}=\int^{T}_{0}l(\Om,\dot\Om)\,dt.
\]

$\bullet$ {\it Discrete setting}: Our goal is to find and algorithm
in $N$ steps of time size $h$, i.e. $Nh=T$, that both minimizes the
cost functional and respects the boundary conditions above. In order
to that, we fix a discretization setting
\begin{equation}\label{DiscreteSetting}
R_{k+1}=R_{k}\,\tau(h\Omega_k),\quad \delta R_{k}=R_{k}\eta_{k},
\end{equation}
where $\eta_k\in\alg$ such that $\eta_{0}=\eta_{N}=0$ and
$\tau(h\Omega_k)\in SO(3)$ is choosen to be a general retraction map. As mentioned before, the first equation $R_{k+1}=R_{k}\,\tau(h\Omega_k)$ is
called the reconstruction equation. From (\ref{DiscreteSetting}), it is easy to obtain the variations of the algebra elements, namely
\begin{equation}\label{deltaOme}
\delta\Omega_k=\mbox{d}\tau^{-1}_{h\Omega_k}(-\eta_k+\Ad_{\tau(h\Omega_k)}\eta_{k+1})/h,
\end{equation}
where $\Ad_g\,\xi=g\,\xi\,g^{-1}$, being $\xi\in\mathfrak{g}$ and $g\in G$.


Our discretization choice enables us to work with algebra
elements instead of group ones. Thus, we define the discrete
function $l_{d}:\so\times\so\Flder\R$ like
$l_{d}(\Om_{k},\Om_{k+1})=hl(\Omega_{k},\frac{\Om_{k+1}-\Om_{k}}{h})$,
where $l(\Om,\dot\Om)$ is explicitly defined in (\ref{LagCont}). We
have set the usual discretization for the derivative
$\dot\Om_{k}=\frac{\Om_{k+1}-\Om_{k}}{h}$. In consequence, let the
{\bf discrete cost functional} be
\begin{equation}\label{DisCostFunctional}
\mathcal{C}_{d}=\sum_{k=0}^{N-1}l_{d}(\Omega_{k},\Omega_{k+1}).
\end{equation}
Therefore, our original optimal control problem defined by $l$ and the boundary conditions $(R(0),\Omega(0))$ and $(R(T),\Omega(T))$ have become a discrete Lagrangian problem with discrete action sum (\ref{DisCostFunctional}). Applying the Hamilton's principle, taking into account the right trivialized derivative of the retraction map defined in (\ref{Retr}) and considering (\ref{deltaOme}), we obtain the discrete equations of motion:
\begin{eqnarray}\nonumber
&&\Ad^*_{\tau(h\Om_{k-1})}(\mbox{d}\tau^{-1}_{h\Omega_{k-1}})^*\lp D_1l_d(\Omega_{k-1},\Omega_k)+D_2l_d(\Omega_{k-2},\Omega_{k-1})\rp\\\nonumber
&&-(\mbox{d}\tau^{-1}_{h\Omega_{k}})^*\lp D_1l_d(\Omega_{k},\Omega_{k+1})+D_2l_d(\Omega_{k-1},\Omega_{k})\rp=0,\\\label{EqsSO}\\\nonumber
&&k=2,...,N-1,
\end{eqnarray}
where $D_1$ and $D_2$ represent the partial derivative w.r.t. the first and second variables respectively.

$\bullet$ {\it Boundary conditions}: from our discretization choice $R_{k+1}=R_k\tau(h\Omega_k)$, is clear that fixing $\Omega_k$ implies constraints in the neighboring points, in this case $R_{k+1}$ and $R_k$. If we allow $\Omega_N$, that means constraints at the points $R_N$ and $R_{N+1}$. Since we only consider time points up to $t=Nh$, having a constraint in the beyond-terminal configuration point $R_{N+1}$ makes no sense. Hence, to ensure that the effect of the terminal constraint on $\Omega$ is correctely accounted for, the set of unknown algebra points ({\it velocities}) must be reduced to $\Omega_{0:N-1}$. Moreover, we can set $\Omega_0=\Omega(0)$, which reduces again, since $\Omega(0)$ is fixed, the unknown velocities to $\Omega_{1:N-1}$.

On the other hand, the boundary condition $R(T)$ is enforced by the relation $\tau^{-1}(R_N^{-1}R(T))=0$. Recalling that $\tau(0)=e$, this last expression just means that $R_N=R(T)$. Moreover, it is possible to translate it in terms of $\Omega_k$ such that there is no need to optimize over any of the configurations $R_k$. In that sense, (\ref{EqsSO}) together with
\[
\tau^{-1}\lp\tau(h\Omega_{N-1})^{-1}...\tau(h\Omega_0)^{-1}R_0^{-1}R(T)\rp=0,
\]
form a set of $3(N-1)$ equations (since dim $\lp\mathfrak{so}(3)\rp=3$) for the $3(N-1)$ unknowns $\Omega_{1:N-1}$. Consequently, the optimal control problem has become a nonlinear root finding problem. From the set of velocities $\Omega_{0:N-1}$ and boundary conditions $(R(0), R(T))$, we are able to reconstruct the configuration trajectory by means of the reconstruction equation $R_{k+1}=R_k\tau(h\Omega_k)$.

$\bullet$ {\it Cayley map}: the group of rigid body rotations is represented by $3\times 3$ matrices with orthonormal column vectors corresponding to the axes of a right-handed frame attached to the body. On the other hand, the algebra $\alg$ is the set of $3\times 3$ antisymmetric matrices.
A $\alg$ basis can be constructed as $\lc\hat e_{1},\hat e_{2},\hat e_{3}\rc$, $\hat e_{i}\in\alg$, where $\lc e_{1},e_{2},e_{3}\rc$ is the standard basis for $\R^{3}$. Elements $\xi\in\alg$ can be identified with the vector $\omega\in\R^{3}$ through $\xi=\omega^{\alpha}\,\hat e_{\alpha}$, or $\xi=\hat\omega$. Under such identification the Lie bracket coincides with the standard cross product, i.e., $\ad_{\hat\omega}\,\hat\rho=\omega\times\rho$, for some $\rho\in\R^{3}$. Using this identification and recalling the hat isomorphism $\hat\cdot$ defined above, we have
\begin{equation}\label{caySO}
\ca(\hat\omega)=I_{3}+\frac{4}{4+\parallel\omega\parallel^{2}}\lp\hat\omega+\frac{\hat\omega^{2}}{2}\rp,
\end{equation}
where $I_{3}$ is the $3\times 3$ identity. The linear maps $\mbox{d}\tau_{\xi}$ and $\mbox{d}\tau_{\xi}^{-1}$ are expressed as the $3\times 3$ matrices
\begin{equation}\label{Dtau}
\mbox{d}\ca_{\omega}=\frac{2}{4+\parallel\omega\parallel^{2}}(2I_{3}+\hat\omega),\,\,\,\,\,\mbox{d}\ca_{\omega}^{-1}=I_{3}-\frac{\hat\omega}{2}+\frac{\omega\,\omega^{T}}{4}.
\end{equation}

\subsection{Example: Cosserat rod}

This example is also known as Kirchhoff's rod. The Cosserat theory
of rods is given in the Lagrangian setting. A static rod corresponds
to a Lagrangian system where the energy density takes the role of
the Lagrangian function.

The potential energy density is the object of most importance in rod
theory. This energy density function (depending on the space curve
parameter) is equivalent to the Lagrangian function of a
time-dependent mechanical system, such that the static equilibrium
equations of a rod correspond to the Euler-Lagrange equations of the
latter.

In this subsection we develop a discrete theory for the Cosserat rod
and treat the associated optimal control problem. An alternatively
formulation of the discrete theory for  the study of symmetries is
given in \cite{julelior}.

The original problem is defined on the tangent bundle of the
manifold $Q=SO(3)\times\R^{3}=SE(3)$ by means of the potential
energy $\pot=\potint+\potext:TQ\Flder\R$. The variables of our
problem are $(R, r, \dot R,\dot r)$, where both $r,\dot r\in\R^{3}$,
$R\in SO(3)$ and $\dot R\in T_{R}SO(3)$. If we assume that the
$\potint$ is frame independent then
\[
\potint(R,r,\dot R,\dot r)=\potbar(R^{-1}\dot R,R^{-1}\dot
r)=\potbar(u,v),
\]
where $\hat u=R^{-1}\dot R\in\so$ and $v=R^{-1}\dot r\in\R^{3}$.
Therefore, our new problem is defined in the left-trivialized
tangent space  $SE(3)\times\se$ as $\pot=\potbar(u,v)+\potext(R,r)$.
With some abuse of notation, let define he elements of $SE(3)$ and $\se=\so\times\R^{3}$  as
\begin{equation}\label{def}
\Phi=(R,r)=\lp\begin{array}{cc}
R&r\\
0_3&1
\end{array}\rp\in SE(3),\,\,\,\phi=(u,v)=\lp\begin{array}{cc}
\hat u&v\\
0_3&0
\end{array}\rp\in\se,
\end{equation}
where $0_3$ is the null $1\times 3$ matrix (both $\Phi$ and $\phi$ are $4\times 4$ matrices). Finally, the total potential
energy is
\[
V=\int^{T}_{0}[\potbar(u,v)+\potext(R,r)]\,dt.
\]
The equilibrium configurations of any static system coincide with
the critical points of the potential energy. In order to obtain the
equations of motion, we consider the following
\begin{equation}\label{deltas}
\delta\hat u=[\hat u,\hat\Su]+ \frac{d}{dt}\hat\Su,\quad \delta v=\hat u\Sv-\hat\Su v+\frac{d}{dt}\Sv,
\end{equation}
where
\begin{equation}\label{Sigmas}
\hat\Su=R^{-1}\delta R\in\so,\,\,\,\,\,\,\,\,\Sv=R^{-1}\delta r\in\R^{3}
\end{equation}
are independent and satisfy the boundary conditions
$\Su(0)=\Su(T)=\Sv(0)=\Sv(T)=0$. It is easy to imagine that both elements form a point in $\se$, namely
\[
\Sigma=\lp\begin{array}{cc}
\hat\Su&\Sv\\
0_3&0
\end{array}\rp.
\]
Taking variations of $V$,
considering equations (\ref{deltas}) and the redefinition
\begin{equation}\label{n,m}
n=\frac{\der\potbar(u,v)}{\der v},\,\,\,\,\,\,\,\,\
m=\frac{\der\potbar(u,v)}{\der u}
\end{equation}
and
\begin{equation}\label{f,l}
f=\frac{\der\potext(R,r)}{\der r}\qquad
l=\frac{\der\potext(R,r)}{\der R},
\end{equation}
which we consider the control forces, we finally arrive to the
equations of motion
\begin{eqnarray}\nonumber
&&\dot n+n\times u+f=0,\\\label{CrosEquations} &&\dot m+n\times
v+m\times u+l=0.
\end{eqnarray}
For more details see \cite{julelior}

The optimal control problem consists on finding a trajectory of the
state variables and control inputs that minimize the cost functional
\[
\mathcal{C}=\int_{0}^{T}\lp f^{2}+\rho_{1}^{2}l^{2}\rp dt,
\]
where $\rho_{1}$ is a weight constant. The control problem is
subject to the following boundary conditions $\Phi(0)=(R(0),
r(0))$, $\phi(0)= (u(0),v(0))$ and $\Phi(T)=(R(T),r(T))$, $\phi(T)=(u(T),v(T))$ belonging
to $SE(3)\times\se$.

As in the rigid body example, from eqs. (\ref{CrosEquations}) we can
obtain an expression of $f$ and $l$ in terms of the other variables.
Furthermore, differentiating equations (\ref{n,m}) with respect to
time, we can find out $\dot n$ and $\dot m$ in terms of $\lp(u,v),(\dot u,\dot v)\rp$ if we assume $\potbar(u, v)$ twice differentiable, i.e.,
$\lp\begin{array}{c}
\dot n\\
\dot m
\end{array}
\rp=\mathcal{H}(u,v)\lp\begin{array}{c}
\dot u\\
\dot v
\end{array}
\rp$, where $\mathcal{H}$ is the Hessian matrix of $\potbar(u,v)$.
Now, setting the function $L:\se\times\se\Flder\R$ as $L((u,v),(\dot u,\dot v))=\left[ f((u,v),(\dot u,\dot v))\right]^{2}+\rho_{1}^{2}\left[l((u,v),(\dot u,\dot v))\right]^{2}$, our problem reduces to extremize the control
functional
\begin{equation}\label{Cost}
\mathcal{C}=\int_{0}^{T}L((u,v),(\dot u,\dot v))\,dt=\int_0^T L(\phi,\dot\phi)\,dt,
\end{equation}
subject to the boundary conditions above. For sake of completeness
we can write down the explicit form of $L$, namely
\begin{eqnarray*}
&&L((u,v),(\dot u,\dot v))=f((u,v),(\dot u,\dot v))^{2}+\rho_{1}^{2}l((u,v),(\dot u,\dot v))^{2}=\\
&&\lp\mathcal{H}_{11}(u,v)\,\dot u+\mathcal{H}_{12}(u,v)\,\dot v+\der_{v}\potbar(u,v)\times u\rp^{2}+\\
&&+\rho_{1}^{2}(\mathcal{H}_{21}(u,v)\,\dot u+\mathcal{H}_{22}(u,v)\,\dot v+\\
&&+\der_{v}\potbar(u,v)\times v+\der_{u}\potbar(u,v)\times u)^{2}.
\end{eqnarray*}

$\bullet$ {\it Discrete Setting}: again we look for an algorithm
minimizing the cost functional (\ref{Cost}) and subject to the
boundary conditions. Firstly, we define the discrete Lagrangian
function $L_d: {\mathfrak se}(3)\times {\mathfrak se}(3)\longrightarrow \R$ as

\[L_{d}(\phi_k,\phi_{k+1})=hL\lp
\phi_{k},\frac{\phi_{k+1}-\phi_{k}}{h}\rp
\]
and
then the {\bf discrete cost functional}
\begin{equation}\label{CostFuncCos}
\mathcal{C}_{d}=\sum_{k=0}^{N-1}L_{d}(\phi_{k},\phi_{k+1}).
\end{equation}
From now on, our discussion is equivalent to the rigid body example developed in (\ref{rb}). We fix the discretization setting
\begin{equation}\label{recc}
\Phi_{k+1}=\Phi_k\tau(h\phi_k),\,\,\,\,\,\delta\,\Phi_k=\Phi_k\Sigma_k,
\end{equation}
where $\Sigma_k\in\se$ s.t. $\Sigma_0=\Sigma_N=0$ and $\tau:\se\Flder SE(3)$ is a general retraction map. Consequently, the variations of $\phi_k$ are
\[
\delta\phi_k=\mbox{d}\tau^{-1}_{h\phi_k}(-\Sigma_k+\Ad_{\tau(h\phi_k)}\Sigma_{k+1})/h,
\]
and the discrete equations of motion:
\begin{eqnarray*}
&&\Ad^*_{\tau(h\phi_{k-1})}(\mbox{d}\tau^{-1}_{h\phi_{k-1}})^*\lp D_1L_d(\phi_{k-1},\phi_k)+D_2L_d(\phi_{k-2},\phi_{k-1})\rp\\
&&-(\mbox{d}\tau^{-1}_{h\phi_{k}})^*\lp D_1L_d(\phi_{k},\phi_{k+1})+D_2L_d(\phi_{k-1},\phi_{k})\rp=0,\\\\
&&k=2,...,N-1,
\end{eqnarray*}

$\bullet$ {\it Boundary conditions}: our reconstruction equation $\Phi_{k+1}=\Phi_k\tau(h\phi_k)$ and boundary conditions $(\Phi(0),\phi(0))$, $(\Phi(0),\phi(0))$ reduce our set of unknowns to $\phi_{1:N-1}$. The discrete equations of motion together with the boundary condition $\Phi(T)=\Phi_N$ enforced by the equation
\[
\tau^{-1}\lp\tau(h\phi_{N-1})^{-1}...\tau(h\phi_0)^{-1}\Phi_0^{-1}\Phi(T)\rp=0,
\]
where $\Phi^{-1}$ is given by
\[
\Phi^{-1}=\lp\begin{array}{cc}
R^{-1}&-R^{-1}r\\
0_3&1
\end{array}\rp,
\]
form a set of $6(N-1)$ equations for the $6(N-1)$ unknowns $\phi_{1:N-1}$ (since $\mbox{dim}(\se)=6$). Again, the optimal control problem has become a nonlinear root finding problem.

$\bullet$ {\it Cayley map}: considering the elements of $SE(3)$ and $\se$ defined in (\ref{def}), the Cayley transform $\ca: \se\Flder SE(3)$ is defined by

\begin{equation}\label{mat}
\ca(\phi)=\lp\begin{array}{cc}
\ca_{SO(3)}(\hat u)&\mbox{d}\ca_{u}\,v\\
0_3&1
\end{array}
\rp,
\end{equation}
where $\ca_{SO(3)}:\alg\Flder SO(3)$ is given by (\ref{caySO}) and $\mbox{d}\ca:\R^{3}\Flder\R^{3}$ by (\ref{Dtau}).

\subsubsection{A direct computation} Choosing $\tau=\ca$ in (\ref{recc}) and taking into account (\ref{mat}), the reconstruction equation $\Phi_{k+1}=\Phi_k\ca(h\phi_k)$ splits as follows:
\[
R_{k+1}=R_{k}\caso(h\hat u_{k}),\,\,\,\,\,\,\,r_{k+1}=r_{k}+hR_{k}\mbox{d}\ca_{hu_{k}}(v_{k}).
\]
For sake of simplicity, we take a truncation of the second equation
such that the {\bf reconstruction setting} stands as
\begin{equation}\label{recSE3}
R_{k+1}=R_{k}\caso(hu_{k}),\,\,\,\,\,\,\,r_{k+1}=r_{k}+hR_{k}v_{k}.
\end{equation}
The second equation in (\ref{recSE3}) clearly represents the easiest discretization of the frame independence condition $v=R^{-1}r$, which in our opinion makes the truncation non-trivial. In order to complete the discrete setting, we define
$g_{k}=\caso(h\hat u_{k})$ and the variations of the $SE(3)$ elements as
\begin{equation}\label{sur}
\delta R_{k}=R_{k}(\widehat\Su)_{k},\,\,\,\,\,\, \delta
r_{k}=R_{k}(\Sv)_{k},
\end{equation}
such that $\widehat\su_{0}=\widehat\su_{N}=0_{3\times 3}$, $\sv_{0}=\sv_{N}=0$.

By means of (\ref{recSE3}) and (\ref{sur}) we can completely
determine $\delta u_{k}$ and $\delta v_{k}$ in terms of $u_{k}$,
$v_{k}$, $\su_{k}$ and $\sv_{k}$:

\begin{eqnarray*}
&&\delta u_k=\frac{1}{h}\Big[\Ad_{g_k}(\Su)_{k+1}-(\Su)_{k}+\frac{h}{2}\ad_{\hat u_k}(\Su)_k-\frac{h}{2}\ad_{\hat u_k}\Ad_{g_k}(\Su)_k\\
&&+\frac{h^2}{4}\hat u_k(\Su)_k\hat u_k-\frac{h^2}{4}\hat u_k\lp\Ad_{g_k}(\Su)_{k+1}\rp\hat u_k\Big],\\\\
&&\delta v_{k}=-(\widehat\Su)_{k}v_{k}+\frac{1}{h}g_{k}(\Sv)_{k+1}-\frac{1}{h}(\Sv)_{k}.
\end{eqnarray*}
Taking variations of
$\mathcal{C}_{d}$ in (\ref{CostFuncCos}) and after long calculations, we arrive to the following algorithm:
\begin{eqnarray}\nonumber
&&\Ad^{*}_{g_{k-1}}\Upso_{(k-2,k-1,k)}-\Upso_{(k-1,k,k+1)}+\\\nonumber
&&+\frac{h}{2}\ad^{*}_{\hat u_{k}}\Upso_{(k-1,k,k+1)}-\frac{h}{2}\Ad^{*}_{g_{k-1}}\ad^{*}_{g_{k-1}}\Upso_{(k-2,k-1,k)}+\\\nonumber
&&+\frac{h^{2}}{4}\hat u_{k}^{*}\Upso_{(k-1,k,k+1)}\hat u^{*}_{k}-\frac{h^{2}}{4}\Ad^{*}_{g_{k-1}}
\hat u_{k-1}^{*}\Upso_{(k-2,k-1,k)}\hat u^{*}_{k-1}+\\\label{Churro}
&&-h[\Uptres_{(k-1,k,k+1)},\, v_{k}]=0,\\\nonumber
\\\nonumber
&&g_{k-1}^{T}\Uptres_{(k-2,k-1,k)}-\Uptres_{(k-1,k,k+1)}=0,\qquad
k=2,...,N-2.
\end{eqnarray}

\begin{eqnarray}\nonumber
&&R_{k+1}=R_{k}\ca_{SO(3)}(h\hat u_{k}),\hspace{1.6cm}
k=0,...,N-1\\\label{REC2}\\\nonumber
&&r_{k+1}=r_{k}+hR_{k}v_{k},\hspace{2.5cm}k=0,...,N-1.
\end{eqnarray}
Here
$\Upso\in\mathfrak{so}^{*}(3)$ and $\Uptres\in\R^{3}$, stands for
\begin{eqnarray*}
&&\Upso_{(a,b,c)}:=D_{1}L_{d}(u_{b},v_{b}, u_{c},v_{c})+D_{3}L_{d}(u_{a},v_{a},u_{b},v_{b}),\\
&&\Uptres_{(a,b,c)}:=D_{2}L_{d}(u_{b},v_{b},u_{c},v_{c})+D_{4}L_{d}(u_{a},v_{a},u_{b},v_{b}),
\end{eqnarray*}
being $(a,b,c)$ integers from $2$ to $N-2$. Both operators $\Ad^{*}$
and $\ad^{*}$ act over elements of $\dso$. The dual algebra element
$\xi^{*}\omega\xi^{*}\in\dso$ is defined such that
$\bra\xi^{*}\omega\xi^{*},\eta\ket=\bra\omega,\xi\eta\xi\ket$ for
$\omega\in\dso$, $\xi,\eta\in\so$ and $\bra\cdot,\cdot\ket$ the
natural pairing between $\so$ and $\dso$.

Finally, we have obtained an algorithm that approximates in an
implicit and non linear way the solution of the optimal control
problem for the Cosserat rod setting.

\section{Conclusions and Future works}

\subsection{Conclusions}
In this paper, we have designed new variational integrators for
optimal control of mechanical systems showing  how developments in
the theory of discrete mechanics and variational methods
\cite{mawest} can be used to construct numerical optimal control
algorithms with certain desirable features. The methods are
available for developing integrators on higher-order problems. The
main idea is to use discrete variational calculus on Lie groups using 
the Lagrange-d'Alembert principle and to derive the discrete
Euler-Poincar\'e equation for discrete Lagrangians corresponding to
a discretization of the second order Lagrangian defined on the
trivialized space (left-trivialized) $G\times 2\mathfrak{g}$.

It is also possible to use  our techniques and the numeric integrator obtained in this
paper for other interesting problems, like for instance the theory of $k$-splines on $SO(3)$ \cite{FHR2010}, \cite{No}.
In this paper, we show two applications of second-order mechanics on the Lie groups on
 $SO(3)$ and $SE(3),$ the rigid body and the
Cosserat rod, respectively.

\subsection{Future Work}
A complete study of symmetry reduction, discrete hamiltonian
description, preservation of geometric structure and numerical
simulations will be developed in a future paper. This discrete
approach will be studied and adapted to the Lie groupoid setting
\cite{CoMa2011a}, \cite{JiMa}, \cite{groupoid}. 
One interesting point, for future work, is to extend our methods to underactuated constraints systems using discrete constrained variational calculus
(see \cite{CMdD} for the continuous counterpart).  
The case of optimal
control problems for mechanical systems with nonholonomic
constraints will be also studied using some of the ideas exposed along the paper
\cite{nonholo}.
\section{Acknowledgments}
This work has been  supported by MICINN (Spain) Grant
MTM2010-21186-C02-01,  MTM2009-08166-E,  project "Ingenio
Mathematica" (i-MATH) No. CSD 2006-00032 (Consolider-Ingenio 2010)
and IRSES-project "Geomech-246981''. L.Colombo also wants to thank
CSIC and JAE program for a JAE-Pre grant.

\end{document}